\documentstyle{article}
\title{\bf QUANTUM STATE AND SPONTANEOUS SYMMETRY BREAKING IN GRAVITY}
\author{PEDRO F. GONZALEZ-DIAZ.\\
\small {\it Instituto de Matem\'aticas y F\'{\i}sica Fundamental,
Consejo Superior de Investigaciones Cientificas} \normalsize\\
\small {\it Serrano 121, 28006 Madrid, Spain\\} \normalsize
}
\date{}
\begin{document}
\maketitle
\large
\setlength{\baselineskip}{0.5cm}

\vspace{2cm}

\begin{center}
ABSTRACT
\end{center}
\begin{center}
\small
A spontaneous symmetry breaking mechanism is used in quantum gravity to obtain
a convergent\\
positive definite density-matrix as the most general quantum state of Euclidean
wormhole theory.
\normalsize
\end{center}

\vspace{1cm}

The quantum state in gravity is given by a functional
\begin{equation}
\int_{C}\delta g_{\mu\nu}\delta\phi e^{-I[g_{\mu\nu},\phi]},
\end{equation}
where $I$ is the Euclidean action. If the integration is over the class $C$
of Euclidean four-geometries and matter field configurations $\phi$ which
match the prescribed data on three-boundaries that divide the four-manifold
in disconnected parts, the probability (1) factorizes in a product of wave
functions describing pure states. For the most general case where the
three-boundaries do not divide the four-manifold, the class $C$ would
contain Euclidean four-geometries and matter field configurations which
should also
match the data on the orientation reverse of the copy three-boundaries. The
state becomes then a mixed state given by a density matrix.$^{1}$ However,
when the density matrix is computed as a propagator on superspace, it
diverges.$^{1}$ In this report we shall suggest an alternate procedure
to compute a
finite definite density matrix based on spontaneous symmetry breaking.

Consider a theory in which a scalar field $\Phi$ with mass $m$ couples
conformally to Einstein-Hilbert gravity. Restricting to Euclidean
Robertson-Walker geometry with scale factor $a$, we obtain for the action
\begin{equation}
I=\frac{1}{2}\int d\eta
N(\frac{\dot{\chi}^{2}}{N^{2}}+\chi^{2}-\frac{\dot{a}^{2}}{N^{2}}
-a^{2}+m^{2}a^{2}\chi^{2}),
\end{equation}
where the dot means differentiation with respect to conformal time
$\eta=\int\frac{d\tau}{a}$,
$N$ is the lapse function and $\chi=(\frac{4\pi G}{3})^{\frac{1}{2}}a\Phi$. The
equations of motion for $\chi$ and $a$ derived from (2) in the gauge $N=1$ are
$\ddot{\chi}=\chi+m^{2}a^{2}\chi$ and $\ddot{a}=a-m^{2}\chi^{2}a$. We notice
nevertheless that these two equations transform into each other by using the
ansatz $\chi=ia$ and that, therefore, the set of such two equations remains
invariant under this ansatz, such as the Hamiltonian action and constraints do.
The equations of motion can then be written as the two formally independent
expressions $\ddot{\chi}=\chi-m^{2}\chi^{3}$ and $\ddot{a}=a+m^{2}a^{3}$. Yet,
making $\chi$ imaginary is classically equivalent to have a scalar axionic
field, so one should substract that part of the surface term which corresponded
to the independent field alone. For transitions with fixed $\Phi$ we should
require
a constant boundary term $S_{B}=Const.NT$. The value of the constant can be
fixed by demanding consistency of the first-integrated equations of motion for
$a$
and $\chi$
\begin{equation}
\dot{\chi}^{2}=\chi^{2}-\frac{1}{2}m^{2}\chi^{4}+R_{0}^{2};\dot{a}^{2}
+\frac{1}{2}m^{2}a^{4}-R_{0}^{2},
\end{equation}
(where $R_{0}^{2}$ is an integration constant) with the Hamiltonian constraint
\begin{equation}

\end{equation}
It follows that $Const.=R_{0}^{2}$. On the other hand, the ansatz $\chi=ia$
leads to a negative definite action, so in order to make that action free
from conformal divergences, when analytically continuing to Euclidean regime,
 instead of
the usual rotation $t\rightarrow-\tau$ employed to derive (2), one should
Wick-rotate anti-clockwise, $t\rightarrow+i\tau$. This can only be strictly
valid$^{2}$ for constant $\Phi$ which is obviously the case if $\chi=ia$ holds.

Re-expressing $I$ in terms of $\tau\rightarrow-t$ and field $\chi$ alone, we
obtain the lagrangian
\begin{equation}
L(\varphi,a)=-\frac{1}{2}(\varphi^{2}\dot{a}^{2}
+\varphi^{2}a^{2}-\frac{1}{2}m^{2}\varphi^{4}a^{4}+R_{0}^{2}),
\end{equation}
where $\varphi=\frac{\Phi}{m_{p}}$, with $m_{p}$ the Planck mass. The solution
to the equations of motion for the axionic field
$\varphi^{2}=-\varphi_{0}^{2}=$Constant, is
\begin{equation}
a(\tau)=(m\varphi_{0})^{-1}[(1+2m^{2}R_{0}^{2})^{\frac{1}{2}}
\cosh(2^{\frac{1}{2}}m\varphi_{0}\tau)-1]^{\frac{1}{2}},
\end{equation}
which represents a wormhole spacetime. If we re-write (5) as a lagrangian
density
$L(\Phi,a)=m_{p}^{2}L(\varphi,a)a^{-4}$, this looks formally similar to the
lagrangian density for an isotropic and homogeneous Higgs model in Euclidean
time
whenever $\Phi=i(\frac{3}{4\pi G})^{\frac{1}{2}}$ is shifted so that
$\Phi\rightarrow\phi=\Phi+\rho$, and $\frac{1}{3}K=\frac{\dot{a}}{a}$ is
interpreted as
the gauge field $A$. It is easy to see$^{3}$ that $L(\Phi,a)$ is invariant
under the isotropic and homogeneous Euclidean version ($t\rightarrow+i\tau$)
of the Abelian group $U(1)$ of transformations
\begin{equation}
A\rightarrow A-i\dot{\zeta}; \Phi\rightarrow\Phi e^{i\zeta}.
\end{equation}
Note that integration of $A$ leads to $a\rightarrow ae^{-i\zeta}$, so that
$\chi$ and the lagrangian are both invariant under (7) only when the lagrangian
is expressed in terms of $\chi$ alone. Without loss of generality one may
regard that after spontaneous symmetry breaking in our Higgs model it is only
the originally zero real component of the conformal field that acquires some
constant classical part $\rho$, and that Goldstone bosons, which could be
thought of as gravitons arising from spontaneous breakdown of diffeomorphism
invariance$^{4}$, would also appear. Actually, a full Euclidean version of
the usual isotropic and homogeneous Higgs mechanism can be readily obtained.

When $\chi=ia$ holds, the theory (5) has just one zero-energy vacuum, and
the solution to the classical equations of motion is $\bar{a}=0$, with
Euclidean action $S_{E}=0$. Expanding about the classical solution, we get
the usual path integral at the semiclassical limit
\begin{equation}
<0\mid e^{-HT/\hbar}\mid
0>=N[det(-\partial_{\tau}+\omega^{2})]^{-\frac{1}{2}}(1+O(\hbar)),
\end{equation}
where $N$ is a constant and $\omega^{2}=\frac{d^{2}V}{da^{2}}(0)$. For the
ground-state solution to $-\partial_{\tau}^{2}+\omega^{2}$, which corresponds
here to a harmonic oscillator with substracted zero-point energy, (8) becomes
constant, $(\frac{\omega}{\pi\hbar})^{\frac{1}{2}}(1+O(\hbar))$. This would
give a pure quantum state.

Let us now consider $\varphi^{2}$ as the control parameter for the
nonlinear dynamic problem posed by action (2). All the values of $\varphi^{2}$
corresponding to an axionic classical field will be negative and, in the
classical
case, can be continuously varied first to zero (a zero potential critical
point) and then to positive values (the field $\varphi$ has become real, no
longer axionic). In the semiclassical theory this generally is no longer
possible however. Not all values of the squared field $\varphi^{2}$ are then
equally probable. For most cases, large ranges of $\varphi^{2}$-values along
the bifurcation itinerary are strongly suppressed.
Nevertheless, there could still be sudden
reversible quantum jumps$^{5}$ from the negative values to the corresponding
positive values of $\varphi^{2}$. Such jumps would be equivalently expressible
as analytic continuations in $a$ to and from its imaginary values.
The system will first go from the bottom of
potential for $\varphi^{2}<0$ to the sphaleron point of potential for
$\varphi^{2}>0$,
$V(a)=\varphi_{0}^{2}(-\frac{1}{2}a^{2}+\frac{1}{4}m^{2}\varphi_{0}^{2}a^{4})$,
without changing position or energy, and then will be perturbed about the
sphaleron saddle point to fall into the broken vacua where the broken phase
condenses
for a short while, to finally redo all the way back to end up at the bottom
of potential for $\varphi^{2}<0$ again. The whole process may be denoted as a
quantum sphaleron
transition and it is assumed to last a very short time and to occur at a very
low frequency along the large time $T$. Therefore, one can use a dilute
spahaleron approximation.
Thus, for large $T$, besides individual quantum sphalerons, there would be
also approximate solutions consisting of strings of widely separated quantum
sphalerons. In analogy with the instanton case, we shall evaluate the
functional integral by summing over all such configurations, with $n$ quantum
sphalerons centered at Euclidean times $\tau_{1}$,$\tau_{2}$,...,$\tau_{n}$.
In the dilute sphaleron approximation,
one has $(\frac{\omega}{\pi\hbar})^{\frac{1}{2}}(-K_{sph})^{n}(1+O(\hbar))$,
where $K_{sph}$ is an elementary frequency associated to each quantum
sphaleron,
and the sign minus accounts for the feature that particles acquire a negative
energy below the sphaleron barrier.
After
integrating over the locations of the sphaleron centers, the sum over any
number $n$
of sphalerons produces a path integral
\begin{equation}
<0\mid e^{-HT/\hbar}\mid
0>_{sph}=(\frac{\omega}{\pi\hbar})^{\frac{1}{2}}e^{-K_{sph}T}(1+O(\hbar)),
\end{equation}
which is proportional to the semiclassical probability of quantum tunnelling
from $\bar{a}=0$ first to $\bar{a}_{\pm}=\pm(m\varphi_{0})^{-1}$ and then
to $\bar{a
}=0$ again.
Approximation (9) corresponds
to a ground-state energy $E_{0}=\hbar K_{sph}$. Thus, the effect of
the quantum sphalerons should be the creation of an extra nonvanishing
zero-point energy which must correspond to a further level of quantization
which is over
and above that is associated with the usual second quantization of the harmonic
oscillator.

If the sphaleron transitions
are taken into account, then the contribution of
the path integral $<0\mid e^{-HT/\hbar}\mid 0>_{sph}$ will introduce a
time-dependent
factor like (9), with $\omega\sim 1$, in the full quantum state. Now, since
time separation between the initial and final states cannot be known, one
should
integrate over $T$ and the quantum state including wormhole perturbations
becomes (here and hereafter we
omit for the sake of simplicity the small term depending on $O(\hbar)$)
\begin{equation}
\frac{1}{\tau_{p}}\int_{0}^{\infty}dT\Psi[a]\Psi[a']e^{-K_{sph}T},
\end{equation}
where $\tau_{p}$ is the Planck time and
the $\Psi '$s are the wave functions for the initial and final states.
The system may yet undergo further sphaleron-asisted bifurcations,
before completing transition to the bottom of potential (2) within the given
small time interval.
In general, there will be strings formed by $n_{j}$
single transitions consisting of $j$ quantum sphalerons. So, considering
the contributions of all such transition strings leads to the maximal analytic
extension of the wormhole manifold. It can be shown that by summing up all
contributions from
the different strings of $n_{j}$ single quantum sphalerons with $K_{sph}=m_{p}$
in harmonic approximation, we obtain a full density matrix
\begin{equation}
\rho=m_{p}\sum_{j=1}^{\infty}\int_{0}^{\infty}dT\Psi_{j}[a]
\Psi_{j}[a']e^{-jK_{sph}T},
\end{equation}
which gives a convergent density matrix.

\vspace{1cm}

{\bf Acknowledgements}
\vspace{0.3cm}

This work was supported by a C.AICYT Research Project N' 91-0052.

\vspace{1cm}

\noindent\section*{References}
\begin{description}

\item [1] D.N. Page, in {\it Proc. of the Fifth Seminar on Quantum Gravity},
eds.
M.A. Markov, V.A. Verezin and V.P. Frolov (World Scientific, Singapore, 1991).

\item [2] A.D. Linde, {\it Inflation and Quantum Cosmology} (Academic Press,
Boston, 1990).

\item [3] P.F. Gonz\'alez-D\'{\i}az, {\it Mod. Phys. Lett.} (in press)

\item [4] S.B. Giddings, {\it Phys. Lett.} B268 (1991) 17.

\item [5] P.F. Gonz\'alez-D\'{\i}az, {\it Phys. Lett.} B307 (1993) 362.

\end{description}

\end{document}